\documentclass[aps,prd,reprint,nofootinbib,showkeys,showpacs,superscriptaddress,longbibliography,floatfix]{revtex4-1}

\pdfoutput=1

\usepackage{amsmath}
\usepackage{amssymb}
\usepackage[american]{babel}
\usepackage{booktabs}
\usepackage{dcolumn}  
\usepackage[T1]{fontenc}  
\usepackage{graphicx}  
\usepackage[colorlinks]{hyperref}  
\usepackage[utf8]{inputenc}  
\usepackage{multirow}
\usepackage{txfonts}  
\usepackage{url}
\usepackage[dvipsnames]{xcolor}
\usepackage{float}
\usepackage{lipsum}
\newcolumntype{d}[1]{D{.}{.}{#1}}
\newcolumntype{v}[1]{D{,}{,\ }{#1}}


\hypersetup{
	citecolor = cyan,
	linkcolor = red,
	urlcolor = magenta
}

\begin{document}
	
	\title{On a possible cosmological evolution of  galaxy cluster $Y_{\rm X}-Y_{\rm SZE}$ scaling relation}

	\author{R. F. L. Holanda}
	\email{holandarfl@fisica.ufrn.br}
	\affiliation{Universidade Federal do Rio Grande do Norte,
		Departamento de F\'{\i}sica, Natal - Rio Grande do Norte, 59072-970, Brasil}

	\author{W. J. C. da Silva}
	\email{williamjouse@fisica.ufrn.br}
	\affiliation{Universidade Federal do Rio Grande do Norte, Departamento de F\'{\i}sica, Natal - Rio Grande do Norte, 59072-970, Brasil}

	\pacs{}
	
	\date{\today}

	\begin{abstract}
		An important result from self-similar models that describe the process of galaxy cluster formation is the simple scaling relation $Y_{\rm SZE}D_{\rm A}^{2}/C_{\rm XSZE}Y_{\rm X}= C$. In this ratio,  $Y_{\rm SZE}$ is the integrated Sunyaev-Zel'dovich effect flux of a cluster, its x-ray counterpart is $Y_{\rm X}$,  $C_{\rm XSZE}$ and $C$ are constants and $D_{\rm A}$ is the angular diameter distance to the cluster. In this paper, we consider the cosmic distance duality relation validity jointly with type Ia supernovae observations plus $61$ $Y_{\rm SZE}D_{\rm A}^{2}/C_{\rm XSZE}Y_{\rm X}$  measurements as reported by the Planck Collaboration to explore if this relation is  constant in the redshift range considered ($z<0.5$). No one specific cosmological model is used. As basic result, although the data sets are compatible with no redshift evolution within 2$\sigma$ c.l., a Bayesian analysis indicates that other $C(z)$ functions analyzed in this work cannot be discarded. It is worth to stress that the observational determination of an universal $C(z)$ function turns the $Y_{\rm SZE}D_{\rm A}^{2}/C_{\rm XSZE}Y_{\rm X}$ ratio in an useful cosmological tool to determine cosmological parameters.\end{abstract}
	
	\maketitle
	
	\section{Introduction}\label{sec:intro}
	
	Galaxy clusters are the largest virialized astronomical structures in the Universe and, their observations provide   powerful tools to probe its evolution  at $z<2$ \citep{Allen2011}. Moreover, important cosmological information can also be obtained from observations of their physical properties. For instance, from the evolution of galaxy clusters x-ray temperatures and their x-ray luminosity function \citep{Ikebe2002,Mantz2008,Vanderlinde2010} the matter density, $\Omega_{\rm m}$, and the normalization of the density fluctuation power spectrum, $\sigma_8$, can be estimated. Galaxy cluster abundance  as a function of mass and redshift can impose competitive limits with other techniques on the evolution of $\omega$, the dark energy equation-of-state parameter \citep{Albrecht2006,Basilakos2010,Devi2014}. By considering  that the gas mass fraction does not evolve with redshift, x-ray observations of these structures are used as standard rulers in order to constrain cosmological parameters \citep{Sasaki1996,Pen1997,Ettori2003,Allen2007,Gonalves2011,Mantz2014,Holanda2019}. The combination of the x-ray emission of the intracluster medium  with the Sunyaev-Zeldovich  effect (SZE) provides estimates of the angular diameter distance to the cluster redshift \citep{Reese2002,DeFilippis2005, Bonamente2006,Holanda2012}. Multiple redshift image systems behind  galaxy clusters  are used to estimate cosmological parameters via strong gravitational lensing \citep{Lubini2013,Magaa2015}. Cosmographic parameters from  measurements of galaxy clusters distances based on their SZE and x-ray observations also can be performed \citep{Holanda2013}.
	
	On the other hand,  if gravity has the dominant effect in galaxy cluster formation process, it is expected to exist self-similar models that predict simple scaling relations between the total mass and basic galaxy cluster properties \citep{Kaiser1986}, such as x-ray luminosity-temperature, mass-temperature, and luminosity-mass relations. It should be stressed these relations are valid only if the condition of hydrostatic equilibrium holds \citep{Giodini2013}. This assumption also breaks down in disturbed systems undergoing mergers \citep{Poole2007}.  Moreover, the very central core of a galaxy cluster can be out of equilibrium when there is AGN activity. In general, these relations are described as power laws and are a  tool for cosmological analyses, and to study the thermodynamics of intra-cluster medium, then, a departure from this prediction can be used to quantify the importance of non-gravitational processes. Uncertainty in the mass-observable scaling relations is currently the limiting factor for galaxy cluster-based cosmology \citep{Giodini2013,Dietrich2018,Singh2020}.
	
	Another scaling relation useful is that one between the integrated SZE flux ($Y_{\rm SZE}\propto T_{\rm g} M_{\rm g}$), which is an ideal proxy for the mass of the gas in a galaxy cluster, and its counterpart in x-ray ($Y_{\rm X}=M_{\rm gas}T_{\rm g}$): $Y_{\rm SZE}D_{\rm A}^{2}/C_{\rm XSZE}Y_{\rm X} = C$, where $C_{\rm XSZE}$ is a constant related with fundamental quantities\footnote{In this expression: $C_{\rm XSZE} = \frac{\sigma_{\rm T}}{m_{\rm e}c^2} \frac{1}{\mu_{\rm e}m_{\rm p}} \approx 1.416 \times 10^{-19} \frac{{\rm{Mpc}}^2}{M_{\odot}{\rm{keV}}}$.} and $C$ is an arbitrary constant \citep{LaRoque2006,Ferramacho2011,Giodini2013,Mantz2016a,Mantz2016b,Truong2017}. The SZE is independent of redshift, and in contrast to x-ray and optical measurements, it does not undergo surface brightness dimming \citep{Birkinshaw1999}. The comparison between $Y_{\rm SZE}$ and ${Y_{\rm X}}$ provides information on the intracluster medium inner structure and especially the clumpiness. Since  $Y_{\rm SZE}D_{\rm A}^2$ and $Y_{\rm X}$ are expected to scale in the same way with mass and redshift, the  ratio between them is expected to be  constant with redshift \citep{Stanek2010,Fabjan2011,Borgani2011,Giodini2013}. Particularly, if galaxy clusters are isothermal this ratio is exactly equal to unity. In  very recent papers,  a new expression for this ratio was derived for the case where there is a departure from the cosmic distance duality relation (CDDR) validity and/or a variation of the fine structure constant $\alpha$ \citep{Galli2013,Colaco2019}.

	By assuming the flat $\Lambda$CDM cosmology to obtain the angular diameter distance ($D_{\rm A}$) for galaxy clusters, the Planck Collaboration presented SZE and x-ray data (XMM-Newton) from a sample of $61$ local  galaxy clusters ($z \le 0.5$). After bias correction, the  $Y_{\rm SZE}D_{\rm A}^{2}/C_{\rm XSZE}Y_{\rm X}$ relation was completely consistent with the expected slope of
	unity (self-similar context) and the  $Y_{\rm SZE}D_{\rm A}^{2}/C_{\rm XSZE}Y_{\rm X}$ ratio was found $0.95\pm 0.04$. On the other hand, by using Compton Y measurements from the Planck measurements, the Ref. \citep{Mantz2016b} found that the slope of this scaling relation departs significantly from self-similarity model.  \citet{Rozo2012} constrained the amplitude, slope, and scatter of this scaling relation by using SZE data from Planck and x-ray data from the Chandra satellite. It was found a $Y_{\rm SZE}D_{\rm A}^{2}/C_{\rm XSZE}Y_{\rm X}$ ratio of $0.82 \pm 0.024$. The slope of the relation was found to be $\alpha = 0.916 \pm 0.032$, consistent with unity only at $\approx 2.3 \sigma$, and no evidence that the scaling relation depends on cluster dynamical state was verified. \citet{Bender2016} considered three subsamples of the $42$ APEX-SZ clusters and  found that the  power laws for the ${Y_{\rm SZE}}-{Y_{\rm X}}$, ${Y_{\rm SZE}}$-$M_{\rm gas}$ and, $Y_{\rm SZE}$-$T_{\rm X}$ relations have exponents consistent with those predicted by the self similar model, where cluster evolution is dominated by gravitational processes. \citet{Henden2018} studied the redshift evolution of the x-ray and SZE scaling relations for galaxy groups and clusters ($z \approx 1$) in the FABLE suite of cosmological hydrodynamical simulations, while the slopes were verified to be approximately independent of redshift, the normalisations evolved  positively with respect to self-similarity.
	
	Recently, the possibility of determining the value of the Hubble constant by using $Y_{\rm SZE}D_{\rm A}^2$ and $Y_{\rm X}$ observations of galaxy clusters was performed by \citet{Kozmanyan2019}, where hydrodynamic simulations in a specific flat $\Lambda$CDM framework was considered to obtain the $C$ value. By applying the method to a sample of $61$ galaxy clusters, they found $H_0 = 67 \pm 3 \, \rm{km} \, \rm{s}^{-1} \, \rm{Mpc}^{-1}$ in a flat $\Lambda$CDM model. As one may see, the ratio $Y_{\rm SZE}D_{\rm A}^{2}/C_{\rm XSZE}Y_{\rm X}$ can be used as an galaxy cluster angular diameter distance estimator if the quantity $C(z)$ is obtained. However, its value must to be found preferably without using any cosmological model.  Actually, the scaling relations are expected to be calibrated by measuring the mass from weak-lensing analyses  \cite{Marrone2012,Hoekstra2015,Sereno2015a,Mantz2016a}. However, this procedure will likely enlarge the scatter of the relations \cite{Sereno2015b} since the lensing mass of single clusters can be both underestimated and overestimated by a considerable amount \cite{Meneghetti2010,Becker2011,Rasia2012}.
	
	In this work, we consider a deformed scaling relation, such as, $Y_{\rm SZE}D_{\rm A}^{2}/C_{\rm XSZE}Y_{\rm X}= C(z) $, to obtain observational limits on  $C(z)$ functions in the redshift considered ($z<0.5$).  $61$ $Y_{\rm SZE}D_{\rm A}^{2}/C_{\rm XSZE}Y_{\rm X}$ measurements from galaxy clusters as reported by the Planck collaboration, type Ia supernovae  observations and the cosmic distance duality  validity ($D_{\rm A}=D_{\rm L}(1+z)^{-2}$) are considered.\footnote{This kind of approach also was performed in \citet{Holanda2017} to put constraints on a possible evolution of mass density power-law index in strong gravitational lensing. In the same way, \citet{Holanda2017b} used the cosmic distance duality relation validity to put limits  on the gas depletion factor in galaxy clusters and  on galaxy cluster structure \cite{Holanda2015}.} The $C(z)$ functions explored here are: $C_0$, $C_0+C_1\ln (1+z)$, $C_0 + C_1 z$, $C_0(1+z)^{C_1}$ and  $C_0+C_1 z/(1+z)$. We obtain that although the data sets are compatible with a constant  $C(z)$ function within 2 $\sigma$ c.l., a Bayesian analysis indicates that other $C(z)$ functions analyzed cannot be discarded. 
	
	The paper is organized as follows: in the section \ref{sec:method} we describe the method developed to probe $C(z)$, and the data sets used for the purpose; in the section \ref{sec:bayesian} we summarize the Bayesian analysis used in this work; while in section \ref{sec:analysis} contains the analyses and discussions; and finally, in session \ref{sec:conclusions}, the conclusions of this paper.
	
	\section{The $Y_{\rm SZE}D_{\rm A}^{2}/C_{\rm XSZE}Y_{\rm X}$ scaling relation and the Cosmic distance duality relation}\label{sec:method}
	
	As commented earlier, the main aim of this work is put observational constraints on a possible redshift evolution of the ratio  $Y_{\rm SZE}D_{\rm A}^{2}/C_{\rm XSZE}Y_{\rm X}= C(z)$ for a galaxy cluster sample. However, as one may see, the angular diameter distance, $D_{\rm A}$, for each galaxy cluster in the sample will be required to perform our method. Previous works obtained this quantity by considering a specific cosmological model (in general the flat $\Lambda$CDM). Here, we follow another route, i.e., observational constraints on the $C(z)$ functions are obtained by measuring the angular diameter distance from the type Ia supernovae (SNe Ia) observations and CDDR validity, $D_{\rm A}=D_{\rm L}(1+z)^{-2}$. 
	
	The CDDR is the astronomical version of the reciprocity theorem proved long ago \citep{Etherington1933,Ellis2006} and its validity requires only that source and observer are connected by null geodesics in a Riemannian spacetime and that the photons number is conserved. Briefly, if the gravity is a  metric theory and  if the Maxwell equations are valid  the distance duality relation is satisfied \citep{Ellis2006}. In the last years, analyses using observational data have been performed in order to establish whether or not the CDDR holds in practice and no significant departure has been measured (see e.g. \citet{Holanda2016a} and references therein). Then, by combining $Y_{\rm SZE}D_{\rm A}^{2}/C_{\rm XSZE}Y_{\rm X}$ and the CDDR, one may obtain:
	
	\begin{equation}\label{eq:scalling-eq}
	\frac{Y_{\rm SZE}D_{\rm L}^{2}}{(1+z)^4C_{\rm XSZE}Y_{\rm X}}=   C(z).
	\end{equation}
	However,  the $Y_X$ quantity is proportional to $ M_{\rm g}$ ($Y_{\rm X}=T_{\rm X} M_{\rm g}$), which depends on the galaxy cluster  distance such as: $M_g \propto D_{\rm L} D_{\rm A}^{3/2}$ (see details in \citep{Sasaki1996}). This measurement is determined by considering a fiducial ($F$) flat $\Lambda$CDM model, with $\Omega_{\rm m } = 0.3$ and $H_0= 70 \, \rm{km} \, \rm{s}^{-1} \, \rm{Mpc}^{-1}$, which  results in: $M_{\rm g} \propto D_{\rm AF}^{5/2}$. Then, we multiply the $Y_{\rm X}$ quantity by $D_{\rm A}^{5/2}/D_{\rm AF}^{5/2}$  in order to eliminate the dependence of $M_{\rm g}$ with respect to the fiducial model. Thus, the Eq. (\ref{eq:scalling-eq}) becomes:
	
	\begin{equation}\label{cz}
	\frac{Y_{\rm SZE}D_{\rm AF}^{5/2}(1+z)}{D_{\rm L}^{1/2}C_{\rm XSZE}Y_{\rm X}}= C(z).
	\end{equation}
	Then, if one has SNe Ia luminosity distances in the redshifts of a galaxy cluster sample, it is possible to impose observational constraints on the $C(z)$ functions.

	\begin{figure*}[htb]
		\centering
		\includegraphics[width=8.5cm]{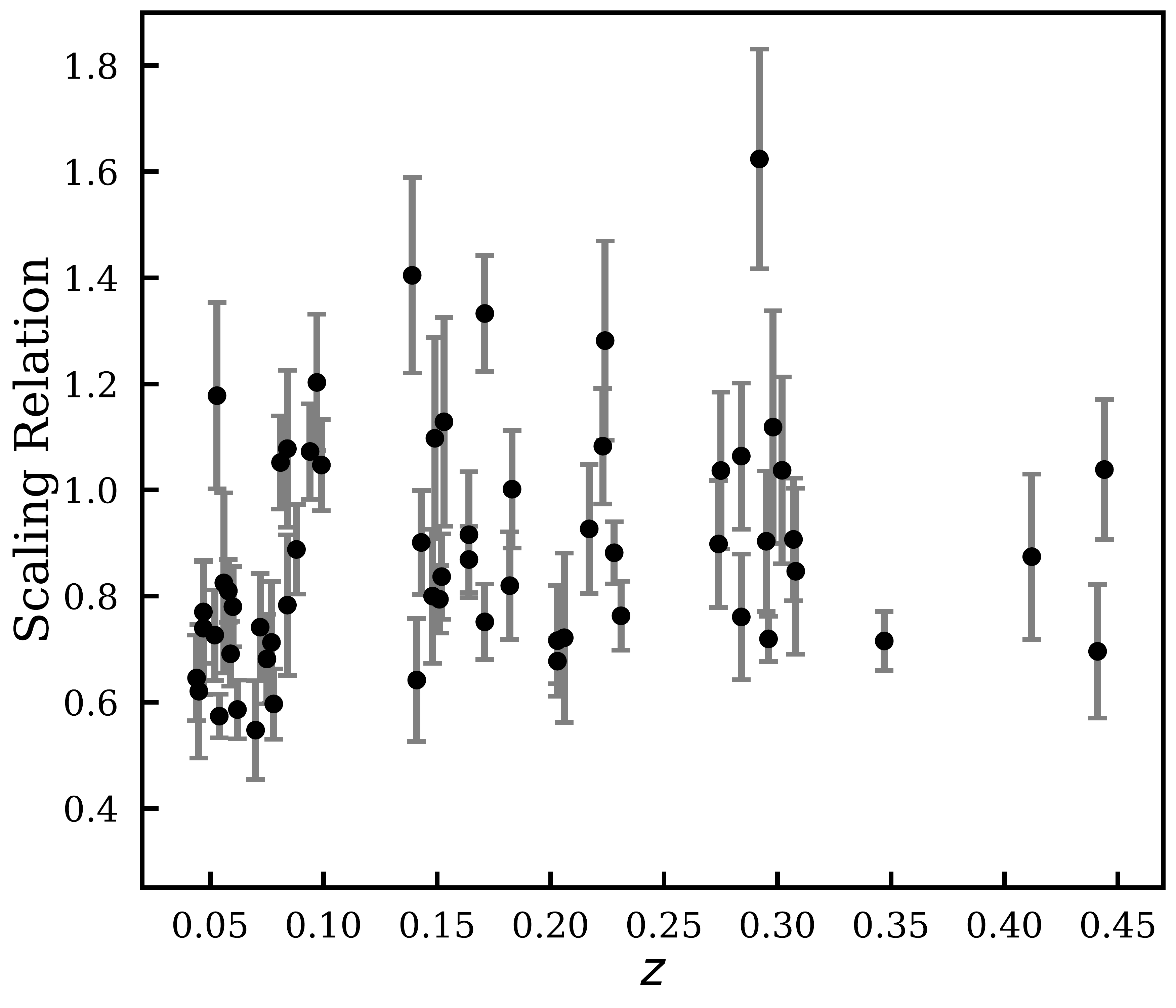}
		\includegraphics[width=8.5cm]{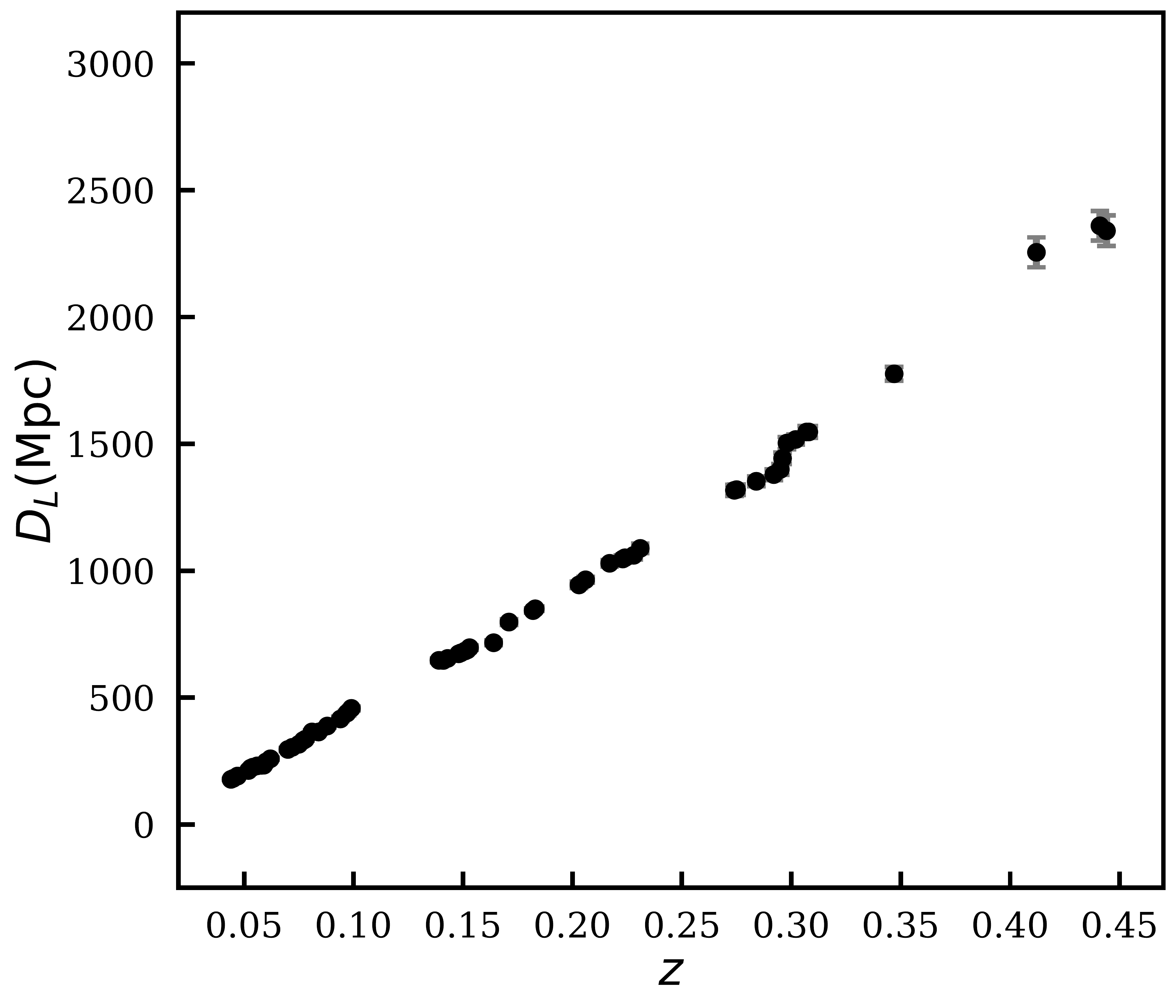}
		\caption{\label{fig1} The left and right figures show the galaxy cluster scaling relation and SNe Ia data, respectively. }
	\end{figure*}

	\subsection{Data}\label{sec:data}
	
	Our  analyses are performed by considering the following samples:
	
	\begin{itemize}
		\item Galaxy clusters: we use $Y_{\rm SZE}-Y_{\rm X}$  measurements of $61$ galaxy clusters obtained from the first {\it Planck mission} \citep{Planck2011} all-sky data set jointly with deep XMM-Newton archive observations.  The galaxy clusters were detected at high  signal-to-noise  within the following redshift interval and mass, respectively: $0.044 \leq z \leq 0.444$ and $2 \times 10^{14} M_{\odot} \leq M_{500} \leq 2 \times 10^{15} M_{\odot}$, where $M_{500}$ is the total mass within  the radius $R_{500}$ that corresponds to the radius that encloses a mass with mean density equal to $500$ times the cosmological critical density $\rho_{\rm c}(z)$ at the redshift of the cluster. The quantities $Y_{\rm X}$ and $Y_{\rm SZE}$ were determined within the $R_{500}$.  The thermal pressure ($P$) of the intra-cluster medium for each galaxy cluster used in our analyses was modeled by \citet{Planck2011} via the universal pressure profile discussed by \citet{Arnaud2010}. This universal profile was obtained by comparing simulated data with observational data (a representative sample of nearby clusters covering the mass range $10^{14}M_{\odot} < M_{500} < 10^{15} M_{\odot}$). The $T_{\rm X}$ quantity was measured in the $[0.15-0.75] R_{500}$ region. The clusters are not contaminated by flares and their morphology are regular enough that spherical symmetry can be assumed (see Fig. \ref{fig1} - left). The galaxy cluster data plotted were obtained by using the Eq. (\ref{cz}) and the luminosity distances as given below.
		\item SNe Ia: we use a sub-sample of the latest and largest Pantheon Type Ia supernovae (SNe Ia) \citep{Scolnic2018} sample in order to obtain $D_{\rm L}$ of the galaxy clusters. The Pantheon SNe Ia compilation consist of  $1048$  spectroscopically confirmed SNe Ia covering the redshift range $0.01 \leq z \leq 2.3$. To perform our test, we need to use SNe Ia and galaxy clusters in the identical redshifts. Thus, for each galaxy cluster, we select SNe Ia with redshifts obeying the criteria $|z_{\rm GC} - z_{\rm SNe Ia}| \leq 0.005$ and calculate the following weighted average for the SNe Ia data:
		
		\begin{equation}
		\begin{array}{l}
		\bar{\mu}=\frac{\sum\left(\mu_{i}/\sigma^2_{\mu_{i}}\right)}{\sum1/\sigma^2_{\mu_{i}}} ,\hspace{0.5cm}
		\sigma^2_{\bar{\mu}}=\frac{1}{\sum1/\sigma^2_{\mu_{i}}}.
		\end{array}\label{eq:dlsigdl}
		\end{equation}
	\end{itemize}
	We end with $61$ measurements of $\bar{\mu}$ and $\sigma^2_{\bar{\mu}}$. The luminosity distance for each galaxy cluster  is obtained through $D_{\rm L} (z)=10^{(\bar{\mu} (z)-25)/5}$ and $\sigma_{D_{\rm L}}^{2}= \left( \frac{\partial D_{\rm L}}{\partial \bar{\mu}} \right)^2\sigma^2_{\bar{\mu}}$ is the associated error over $D_{\rm L}$  (see Fig. \ref{fig1} - right).
	
	\section{Bayesian Inference}\label{sec:bayesian}
	
	The Bayesian inference is a powerful statistical technique for parameter estimation and model selection extensively used in the study of Cosmology and Astronomy \citep{Hobson2002,Trotta2007,Santos2017,Silva2019,daSilva2019,Cid2019,Kerscher2019,daSilva:2020oxl}. The basis of this theory is the Bayes' Theorem, which updates the probability of an event (or hypothesis), based on prior knowledge of conditions that might be related to the event in the light of newly available data (or information). The Bayes's Theorem relates the posterior distribution $P(\Phi|D, M)$, likelihood $\mathcal{L}(D|\Phi, M)$,  the prior distribution $\pi(\Phi|M)$, and the Bayesian evidence $\mathcal{E}(D|M)$ \citep{Trotta2008}:
	
	\begin{equation}\label{bayes}
	P(\Phi|D, M) = \frac{\mathcal{L}(D|\Phi, M) \pi(\Phi|M)}{\mathcal{E}(D|M)},
	\end{equation}
	where $\Phi$ is the set of parameters, $D$ represents the data and $M$ is the model. 
	
	The Bayesian evidence $\mathcal{E}(D|M)$ constitutes a normalization constant in the context of the parameter constraint, however, it becomes a fundamental element in the Bayesian model comparison approach. So, the Bayesian evidence of a model in the continuous parameter space $\Omega$ can be written as:
	
	\begin{equation}\label{evidence}
	\mathcal{E}(D|M) = \int_\Omega \mathcal{L}(D|\Phi, M) \pi(\Phi|M) d\Phi.
	\end{equation} 
	Therefore, the evidence is the distribution of the observed data marginalized over the set of parameters. The most significant feature in the Bayesian model comparison is associated with the comparison of two models that describe the same data. When comparing two models, $M_i$ versus $M_j$, given a set of data, we use the Bayes' factor defined in terms of the ratio of the evidence of models $M_i$ and $M_j$:
	
	\begin{equation}\label{bayes_factor}
	\mathcal{B}_{ij} = \frac{\mathcal{E}_i}{\mathcal{E}_j},
	\end{equation}
	where $\mathcal{E}_j$ and $\mathcal{E}_i$ are the competing models in which we want to compare. To evaluate either the model has favorable or not evidence, we adopted Jeffreys’ scale to interpret the values of the Bayes' factor \citep{Jeffreys61,Trotta2008}.  This scale interprets the evidence as follows: inconclusive if $|\ln \mathcal{B}_{ij}| < 1$, weak if $1 \leq |\ln \mathcal{B}_{ij}| < 2.5$, moderate if $2.5 \leq |\ln \mathcal{B}_{ij}| < 5$ and strong if $|\ln \mathcal{B}_{ij}| \geq 5$. A negative (positive) value for $\ln\mathcal{B}_{ij}$ indicates that the competing model is disfavoured (supported) with respect to the standard model.
	
	Additionally, we assume that both type Ia supernovae and galaxy clusters data set follow a Gaussian likelihood, such as: 
	
	\begin{equation}\label{likelihood}
	\mathcal{L}(D|\Phi, M) \propto \exp\left[-\frac{\chi^2(D|\Phi, M)}{2}\right],
	\end{equation}
	whose $\chi^2$ reads
	
	\begin{equation}
	\chi^2(D|\Phi, M) = \sum_i \left(\frac{C(z_i) - Y(z_i)}{Y_{\text{err}}}\right)^2,
	\end{equation}
	where $C(z)$ are theoretical values obtained from the functions $C_0$, $C_0 + C_1\ln(1+z)$, $C_0 + C_1 z$, $C_0(1+z)^{C_1}$ and $C_0+C_1 z/(1+z)$ that we will analyze, $Y(z)$ is a vector of the observed values given by equation \ref{cz} and $Y_{\text{err}}$ is the error associated with scaling relation measurements.
	
	To implement the statistical analysis, we consider the public package \textsf{MultiNest} \citep{Feroz2008,Feroz2009,Buchner2014} through the \textsf{PyMultiNest} interface \citep{Buchner2014}. To perform this analysis, we choose uniform priors about the free parameters of the models investigated. These priors are $0.5 \le C_0 \le 1.5$ and $-1 \le C_1 \le 1$. Furthermore, to increase the efficiency in the parameter estimate and, in the evaluation of the evidence, we ran the codes five times with a set of $2000$ live points each one. In this way, the number of the posterior distributions was of the order $\mathcal{O}(10^4)$ in each result. So, we concatenated the posteriors and computed the average of evidence.

	\section{Statistical Analyses and Results}\label{sec:analysis}

	\begin{table*}[t]
		\centering
		\caption{\label{tab:results1} Confidence limits for the parameters using galaxy clusters + SNe Ia. The columns show the constraints on each function whereas the rows show the parameter considered in our analysis. The last rows display the Bayesian evidence, the Bayes’ factor and its interpretation, respectively.}
		\begin{tabular}{lccccc}
			\hline
			& $C_0$     & $C_0 + C_1\ln (1+z)$     & $C_0 + C_1z$ & $C_0(1+z)^{C_1}$ & $C_0 + C_1\frac{z}{1+z}$ \\ \hline
			$C_0$       & $0.926\pm 0.025$ & $0.870 \pm 0.046 $ & $0.877 \pm 0.045$   & $0.877 \pm 0.046$         & $0.866 \pm 0.046$      \\ 
			$C_1$ &  -      & $0.390 \pm 0.275$  &  $0.310 \pm 0.230$     & $0.370 \pm 0.280 $         &   $0.460 \pm 0.300$             \\ 
			$\ln \mathcal{E}$      & $-34.116$ & $-34.164$ & $-34.434 $   & $-34.239$          & $-33.925$               \\ 
			$\ln \mathcal{B}_{ij}$ & 0       & $-0.049$ & $-0.318$     & $-0.124$          & $0.189$                 \\ 
			Interpretation & -       & Inconclusive  & Inconclusive & Inconclusive  & Inconclusive           \\ 
			\hline
		\end{tabular}
		
	\end{table*}

	\begin{table*}[t]
		\centering
		\caption{\label{tab:results2} Confidence limits for the parameters assuming the $\Lambda$CDM fiducial. The columns show the constraints on each function whereas the rows show the parameter considered in our analysis. The last rows display the Bayesian evidence, the Bayes’ factor and its interpretation, respectively.}
		\begin{tabular}{lccccc}
			\hline
			& $C_0$     & $C_0 + C_1\ln (1+z)$     & $C_0 + C_1z$ & $C_0(1+z)^{C_1}$ & $C_0 + C_1\frac{z}{1+z}$ \\ \hline
			$C_0$       & $0.883 \pm 0.022$  & $0.821 \pm 0 .042$ & $0.828 \pm 0.041$    & $0.829 \pm 0.039$        & $0.817 \pm 0.041$           \\ 
			$C_1$ & -       & $0.430 \pm 0.250$ & $0.350 \pm 0.210$   & $0.440 \pm 0.265$   & $0.510\pm 0.270$             \\  
			$\ln \mathcal{E}$       & $-37.329$ & $-36.978$ & $-37.282$   & $-37.055$       & $-36.702$               \\ 
			$\ln \mathcal{B}_{ij}$ & 0       & $0.351$ & $0.046$     & $0.273$           & $0.626$                  \\ 
			Interpretation & -       &Inconclusive  & Inconclusive & Inconclusive  & Inconclusive           \\ \hline
		\end{tabular}
	\end{table*}

	In the Tables \ref{tab:results1} and \ref{tab:results2} are shown the mean, $1\sigma$ and $2\sigma$ errors obtained considering the galaxy clusters and SNe Ia and the $\Lambda$CDM fiducial, respectively. By considering the galaxy clusters and SNe Ia data (Table \ref{tab:results1}), we obtained: $C_0 = 0.926\pm 0.025$ for the first function (constant). For $C_0 + C_1\ln (1+z)$ we obtain: $C_0 = 0.870 \pm 0.046 $ and $C_1=0.390 \pm 0.275$. The values obtained for parameters of the $C_0 + C_1z$ function were: $C_0 = 0.877 \pm 0.045$, and $C_1 = 0.310 \pm 0.230$. Then, within 1$\sigma$ c.l., this $C_1$ values are incompatible  with no redshift evolution of the scaling relation. {When we considered the monomial function}, $C_0(1+z)^{C_1}$, and the results obtained were: $C_0 = 0.829 \pm 0.039$ and $C_1 = 0.440 \pm 0.265 $. Again, a possible evolution with redshift is allowed, at least, within 1$\sigma$ c.l.. Finally, for the last function, we obtain: $C_0 = 0.817 \pm 0.041$ and, $C_1 = 0.510\pm 0.270$, as one may see, a $C(z)$ function evolving with redshift also is compatible, at least, within 1$\sigma$ c.l..
	
	{We also considered} the angular diameter distance for each galaxy cluster obtained from the $\Lambda$CDM fiducial ($H_0 = 67.4\pm 0.5 \, \rm{km} \, \rm{s}^{-1} \, Mpc^{-1}$ and $\Omega_{\rm m} = 0.315 \pm 0.007$), the value estimated for the constant function was $\approx 4 \%$ smaller than obtained by galaxy clusters and SNe Ia data.  The results are in good agreement with those from galaxy clusters plus SNe Ia data.
	
	In this point, a question that arises is: which $C(z)$ function describes the behaviour of the $Y_{\rm {SZE}}D_{A}^{2}/C_{\rm {XSZE}}Y_X$ ratio with the redshift? Then, to deal with that, we performed a Bayesian model comparison analysis between the $C(z)$ functions in terms of the strength of the evidence according to the Jeffreys’ scale. To make this, we estimate the values of the logarithm of the Bayesian evidence ($\ln \mathcal{E}$) and the Bayes' factor ($\ln \mathcal{B}$), Tables \ref{tab:results1} and \ref{tab:results2}. These results were obtained considering the priors defined in the last section and, we assumed the constant function as the reference one. By considering the analysis with the galaxy clusters + SNe Ia and the $C(z)$ functions, we obtained positive values of the Bayes' factor and, according to Jeffreys' scale, these functions have favorable inconclusive evidence. We also perform  a Bayesian model comparison by considering  the analysis with galaxy cluster + $\Lambda$CDM model fiducial (Table \ref{tab:results2}). Again, we obtain that the  $C(z)$ functions with a redshift evolution have inconclusive evidence favored by the data.	From both analyses, we can conclude that a possible redshift evolution is allowed by the present data sets within the redshift range considered. 
	
	It is important to stress that the Ref. \cite{Planck2011} divided the complete sample between cool core and non-cool core clusters. Cool core clusters were considered those with central electronic density ($n_{e,0}$) $ > 4 . 10^{-2}$ $ \rm{cm}^{-3}$. In total, 22/62 clusters in the complete sample were classified as such. {We have also performed} our method considering these two subsamples separately and the results obtained  are in full agreement each other. 
	
	\section{Conclusions}\label{sec:conclusions}

In order to use galaxy clusters as a cosmological probe it is needed to deeply know some their intracluster gas and dark matter properties. Particularly,  scaling relations between the observable properties and the total masses of these structures are key ingredient for analysis that aim to constrain cosmological parameters and so they need to be well-calibrated. Particularly,  the ratio $Y_{\rm SZE}D_{\rm A}^{2}/C_{\rm XSZE}Y_{\rm X}= C$ could  be used as a galaxy cluster angular diameter distance estimator if the quantity $C$ is determined without using any cosmological model.

In this paper, we  assumed the cosmic distance duality relation validity and considered  type Ia supernovae observations plus $61$ galaxy cluster $Y_{\rm SZE}D_{\rm A}^{2}/C_{\rm XSZE}Y_{\rm X}$ measurements as reported by the Planck Collaboration in order to verify if such relation is  constant within the redshift considered ($z<0.5$). No specific cosmological model was used. Five $C(z)$ functions describing a  possible evolution of the $Y_{\rm SZE}D_{\rm A}^{2}/C_{\rm XSZE}Y_{\rm X}$ ratio with redshift were explored, namely: $C_0$, $C_0+C_1 \ln(1+z)$, $C_0 + C_1 z$, $C_0(1+z)^{C_1}$ and $C_0+C_1 z/(1+z)$. We obtained that although the data sets are compatible with a constant scaling relation ($C_1=0$) within 2$\sigma$ c.l. (see Table \ref{tab:results1}), a Bayesian analysis indicated that a possible redshift evolution cannot be discarded with the present data sets. We also performed our analysis by using the Flat $\Lambda$CDM model (Planck collaboration) to determine the angular diameter distances to the clusters and the results had negligible changes (see Table \ref{tab:results2}). Moreover, the cool core and non-cool core clusters  were analyzed separately and the results did not depend on the cluster dynamical state.

Finally, it is worth to comment that when more and larger data sets with smaller statistical and systematic uncertainties become available, the method proposed here (based on the validity of the distance duality relation) should improve on the galaxy cluster scaling relation understanding.
	
\begin{acknowledgements}
	RFLH thanks financial support from Conselho Nacional de Desenvolvimento Cientıfico e Tecnologico (CNPq) (No.428755/2018-6 and 305930/2017-6). 
\end{acknowledgements}
	
	\bibliographystyle{apsrev4-1}
	\bibliography{references}

\end{document}